\documentclass[aps,superscriptaddress,showpacs,nofootinbib,twocolumn,10pt]{revtex4}
\usepackage{amscd}
\usepackage{amsmath}
\usepackage{bm}
\usepackage[dvips]{color}  
\usepackage{graphicx}
\usepackage{eufrak}
\usepackage{eucal}

\usepackage{epsfig}
\begin{document}

\title{Superfluidity in Two-Dimensional Imbalanced Fermi Gases}
\author{Heron Caldas}\email{hcaldas@ufsj.edu.br} 
\author{A. L. Mota} 
\author{R. L. S. Farias} 
\author{L. A. Souza} 
\affiliation{Departamento de Ci\^{e}ncias Naturais, Universidade Federal de 
S\~{a}o Jo\~{a}o del Rei,\\ 36301-160, S\~{a}o Jo\~{a}o del Rei, MG, Brazil}
 
\begin{abstract}

We study the zero temperature ground state of a two-dimensional atomic Fermi gas with chemical potential and population imbalance in the mean-field approximation. All calculations are performed in terms of the two-body binding energy $\epsilon_B$, whose variation allows to investigate the evolution from the BEC to the BCS regimes. By means of analytical and exact expressions we show that, similarly to what is found in three dimensions, at fixed chemical potentials, BCS is the ground state until the critical imbalance $h_c$ after which there is a first-order phase transition to the normal state. We find that $h_c$, the Chandrasekhar-Clogston limit of superfluidity, has the same value as in three dimensional systems.  We show that for a fixed ratio $\epsilon_B/\epsilon_F$, where $\epsilon_F$ is the two-dimensional Fermi energy, as the density imbalance $m$ is increased from zero, the ground state evolves from BCS to phase separation to the normal state. At the critical imbalance $m_c$ phase separation is not supported and the normal phase is energetically preferable. The BCS-BEC crossover is discussed in balanced and imbalanced configurations. Possible pictures of what may be found experimentally in these systems are also shown. We also investigate the necessary conditions for the existence of bound states in the balanced and imbalanced normal phase.

\end{abstract}


\maketitle

\section{Introduction}

In the last few years, great experimental advances have permitted the manipulation of trapped neutral ultracold two-spin-components atomic Fermi gases with tremendous accuracy. The essential technics under full domain are the cooling, trapping, the control of the number of atoms in the sample, and the tunning of the inter-atomic (s-wave) interactions via the application of an external uniform magnetic (Feshbach resonance) field~\cite{Inouye}. This allowed the investigation of the crossover from the Bardeen-Cooper-Schrieffer (BCS) phase of weakly bound Cooper pairs to the Bose-Einstein condensate (BEC) phase of strongly bound diatomic molecules in three-dimensional (3D) trapped Fermi gases~\cite{Exp1,Exp2,Exp3}.

These many-body quantum gases are even more exciting when the two-components have mass, chemical potential or population imbalance, since this situation can be met in many areas of physics, from condensed matter to high-density quark matter. Many exotic phases have been proposed as the ground state of these imbalanced Fermi gases, such as the homogeneous Sarma~\cite{BP1} or breached-pair state~\cite{BP2}, the Fulde-Ferrel-Larkin-Ovchinnikov (FFLO) state with modulated order parameter~\cite{FFLO}, phase separation of superfluid and normal components in real space~\cite{Caldas1,Caldas2}, deformed Fermi surfaces~\cite{Armen}, and magnetized superfluid in the BEC side of the resonance~\cite{Sheehy}. 

Recent experiments in imbalanced atomic Fermi gases confined in 3D harmonic traps have observed that these systems phase separate into a unpolarized superfluid core surrounded by a normal outer region~\cite{Exp4,Exp5,Exp6,Exp66,Exp7}. The observation of the FFLO state is an experimental challenge, since it is predicted to exist only in a narrow window of asymmetry between the particle's chemical potentials.  Indeed recent experiments of Shin et al.~\cite{Exp7} have not observed any evidence of the exotic FFLO state in 3D. However, lower dimensions may favor FFLO due to the nesting of the Fermi surfaces, as in the recent experiments with two spin mixture of ultracold $^{6}Li$ atoms trapped in an array of one dimensional (1D) tubes, where FFLO-like correlations were found for a range of polarizations~\cite{Hulet}.

Thus, another very important parameter under control is the trap geometry, which permits the experimental investigation of fermionic atoms in lower dimensions. A two-dimensional (2D) trapped Fermi gas may be achieved in experiments by flattening magnetic or dipolar confinements~\cite{Exp8}, by trapping atoms in specially designed pancake potentials~\cite{Exp9}, by radio frequency (rf)-induced two-dimensional traps~\cite{Exp10}, by gravito-optical surface traps~\cite{Exp11}. Due to these experimental advances, recent studies have reported various features of 2D balanced Fermi gases: a 2D Fermi gas of atoms has been prepared and directly observed \cite{prl_detect_2d_gas}, the first realization of a strongly interacting 2D Fermi gas of atoms has been reported~\cite{Michael1}, the pairing pseudogap in a 2D Fermi gas in the strong coupling regime has been explored~\cite{Michael2}, the biding energy of fermion pairs was measured along the dimensional crossover from 3D to 2D~\cite{prlzwierlein}, and in Ref.~\cite{baur} the contributions of the effective-range corrections to two-body bound-state energies were considered.

Inspired by these experimental progresses, some recent papers have investigated two-component Fermi gases both in 2D~\cite{Tempere,Duan,Zhuang,Simons,Silva,Du,Wan}, as well as in one-dimensional (1D) Fermi systems as, for instance, in Refs.~\cite{Orso,Drumond1,Drumond2,Guan,Paata}. Besides, spin imbalance has also been investigated in the context of metallic systems as, for example, electron-hole bilayer~\cite{Pieri,Kazuo} and multi-band~\cite{Mucio} systems.

In the imbalanced scenario, much effort has been dedicated to study the Fermi polaron problem in which a single spin-$\downarrow$ atom interacts strongly with a Fermi sea of spin-$\uparrow$ atoms~\cite{zollner,parish1,klawnn,schmidt}. Fermi polarons were observed in a tunable Fermi liquid of ultracold atoms~\cite{Polaron}, and experimental evidence for the polaron-molecule transition was recently found~\cite{nature_khol}.

In this paper we study the zero temperature ($T$) ground state of a 2D atomic Fermi gas with chemical potential and population imbalance in the mean-field approximation. In this analysis we do not consider the Berezinskii-Kosterlitz-Thouless (BKT) corrections~\cite{BKT} to the mean-field results in 2D. BKT corrections to mean-field results are more important close (and below) a certain critical temperature $T_c$, while at $T = 0$, as considered in the present paper, all the vortices and antivortices are bound in molecules, so BKT corrections are not relevant. The (topological) BKT transition has been investigated in 2D balanced~\cite{Sa,Duan2}, as well as imbalanced systems~\cite{Devreese}, by taking into account the phase fluctuation effect. We show that in the grand canonical ensemble the fundamental state is BCS up to a critical value of the chemical potentials asymmetry $h_c$, after which there is a quantum phase transition to the normal state. In the canonical ensemble we study in detail the issue of phase separation. We show that the normal phase is always unstable to phase separation when the density asymmetry $m$ is less than the critical imbalance $m_c$. All calculations are performed in terms of the two-body binding energy $\epsilon_B$, whose variation allows to investigate the evolution from the BEC to the BCS regimes. The possibility of the FFLO state is ignored in this work, and we only consider pairing between atoms with equal and opposite momentum.

The main results of this paper may be summarized as:
\newline 
{\bf I.}  We show that in a zero $T$ imbalanced 2D gas of fermionic atoms, BCS is the ground state up to chemical potentials imbalance $h_c$ at which there is a first-order phase transition to the normal state. We find that $h_c$ is the same as in 3D, namely, the Chandrasekhar-Clogston limit of superfluidity.
\newline
{\bf II.} We demonstrate that, as found theoretically and experimentally in 3D, for a fixed two-body binding energy $\epsilon_B$, as the imbalance $m$ is increased from zero the stable states are BCS, phase separation and normal phase.
\newline
{\bf III.} For balanced systems, the BCS-BEC crossover is governed only by $\epsilon_B$. However, for imbalanced systems, the BCS-BEC crossover will depend on $\epsilon_B$ and the imbalance $m$. For a fix value of $\epsilon_B$, the crossover is possible only for asymmetries supported by phase separation. Above a critical value of the number densities imbalance $m_c$, phase separation is destroyed (in favor of the normal phase) as well as the path to BEC regime.
\newline
{\bf IV.} We investigate the conditions for the existence of bound states in the balanced and imbalanced normal phase. We find that in the balanced normal phase at zero temperature, there will be bound states for any value of the two-body binding energy. For the imbalanced normal configuration, we find that stable (and real) bound states exist when $2\mu+|\epsilon_B|> 2h$, where $h$ is the chemical potential imbalance.
\newline
{\bf V.} As we point out in Section~\ref{comp}, the results we obtain in the Canonical ensemble agree with previous works which employed the Grand Canonical ensemble, as it should be, since the phase diagram should be independent of this choice.

The paper is organized as follows: In Sec.~\ref{Int} we present the model Hamiltonian and derive an analytical expression for the zero temperature thermodynamical potential in the mean-field approximation for the two-component gas of fermionic atoms in 2D. In Sec.~\ref{P} we employ this expression to investigate the stability of the possible phases of 2D imbalanced Fermi gases, both at fixed chemical potentials and densities. We also discuss the BCS-BEC crossover in balanced and imbalanced systems. In Sec.~\ref{bs} we study the necessary conditions for the formation of bound states in the normal phase. In Sec.~\ref{comp} we perform a comparison with solid zero temperature results obtained in the recent literature. Finally, Sec.~\ref{conc} is devoted to the conclusions.

\section{The Model Hamiltonian and the Mean-Field Theory}
\label{Int}

We start by considering a 2D nonrelativistic dilute system of fermionic atoms of mass $M$, with two hyperfine states labeled as $\sigma = \uparrow, \downarrow$. This spin $\uparrow$ and $\downarrow$ mixture can be done with the two lowest hyperfine states of $^{6}{\rm Li}$ atoms, as in the 3D experiments~\cite{Exp5,Exp6,Exp7}. This 2D system may be experimentally realized confining a 3D Fermi gas into a single layer (or a stack of layers) by a tight transverse harmonic oscillator potential $V(z)=\frac{m}{2} \omega_z^2z^2$, where $\omega_z$ is the trapping frequency along the confined direction. For very low temperatures and densities such that $k_B T, \epsilon_F<<\hbar \omega_z$, where $\epsilon_F$ is the Fermi energy, and weak longitudinal trapping, the collisions can be considered to be quasi-2D~\cite{prl_detect_2d_gas}. The single-particle dispersion relations are given by $\xi_k=\frac{\hbar^2 k^2}{2 M}$. Throughout the paper we set $\hbar=1$. Since at low densities (appropriate to describe the ultracold trapped Fermi gases we are interested) the form of the potential is not probed, it can be considered that the atoms interact via a contact interaction which is modeled by the following pairing Hamiltonian:

\begin{equation}
\label{H0}
H=H_{0} + H_{\rm int},
\end{equation}
where

\begin{equation}
\label{H}
H_{0}=\sum_{k,\sigma = \uparrow, \downarrow} \epsilon_k^{\sigma} \psi_{\sigma}^{\dagger}(k) \psi_{\sigma} (k),
\end{equation}
is the kinetic (free) part of $H$ and $H_{\rm int}$ is given below. $\psi_{\sigma}^{\dagger}(k)$ and $\psi_{\sigma}(k)$ in Eq.~(\ref{H}) are the creation and annihilation operators, respectively, for the $\uparrow, \downarrow$ particles. To assure population imbalance in the system, we have introduced different chemical potential for the species $\sigma$ as $\mu_{\sigma}=\mu + \sigma h$, where $\sigma h  \equiv \pm h$. Then, the chemical potential $\mu_{\sigma}$ fix the number densities $n_{\sigma}$ of the different fermions. The new dispersions for the free species $\sigma$, relative to their Fermi energies, are $\epsilon_k^{\sigma} \equiv  \xi_k  -\mu_{\sigma}$. The interaction Hamiltonian is given by

\begin{equation}
H_{int}= g \sum_{k, k'} \psi_{\uparrow}^{\dagger}(k) \psi_{\downarrow}^{\dagger}(-k) \psi_{\downarrow} (-k') \psi_{\uparrow} (k'),
\end{equation}
where the bare coupling constant $g$ is negative, to express the attractive (s-wave) interaction between the spin $\uparrow$ and $\downarrow$ fermionic atoms.

After the mean field (MF) approximation and the subsequent diagonalization of the expression for $H_{MF}$, we arrive at the following expression for thermodynamic potential:

\begin{eqnarray}
\Omega &=&-\frac{\Delta^2}{g}+\int_{k<k_1,~k>k_2} \frac{d^2 k}{(2\pi)^2} (\epsilon_k^b - {\cal E} _k^{\beta})
\nonumber
\\
&+& \int_{k_1}^{k_2} \frac{d^2 k}{(2\pi)^2} \epsilon_k^b
\nonumber
\\
&=&-\frac{\Delta^2}{g}+\int_{k<k_1,~k>k_2} \frac{d^2 k}{(2\pi)^2} (\epsilon_k^+ - E_k)
\nonumber
\\
&+& \int_{k_1}^{k_2} \frac{d^2 k}{(2\pi)^2} \epsilon_k^b,
\label{Omega1}
\end{eqnarray}
where, for simplicity of notation we have labeled $\downarrow=a$, $\uparrow=b$. Here we have defined ${\cal{E}}_k^{a,b}= E_k \pm \epsilon_k^{-}$ are the quasiparticle excitations, with $E_k=\sqrt{ {\epsilon_k^{+}}^2+\Delta^2 }$, $\epsilon_k^{\pm} = \frac {\epsilon_k^a \pm \epsilon_k^b}{2}$ and the constant pairing gap is given by $\Delta = - g \int \frac{d^2 k}{(2\pi)^2} \langle \psi_{\downarrow}^{\dagger}(-k) \psi_{\uparrow}^{\dagger}(k) \rangle = \Delta^*$.

In Eq.~(\ref{Omega1})

\begin{equation}
\label{roots}
k_{1,2}^2 = \frac{k_a^2+k_b^2}{2} \pm \frac{1}{2}\sqrt{(k_b^2-k_a^2)^2-16M^2 \Delta^2}
\end{equation}
are the roots of ${\cal E} _k^{\beta}$, and $k_{\alpha}=\sqrt{2M \mu_{\alpha}}$ is the Fermi momentum of the species ${\alpha=a,b}$. From the equation above one sees that the condition for $\Delta$ such that $k_{1,2}$ is real is $\Delta \leq \frac{k_b^2-k_a^2}{4M}$.

To regulate the ultraviolet divergence associated with the first integral in Eq.~(\ref{Omega1}) we introduce~\cite{Randeria}:

\begin{equation}
\label{reg}
\frac{1}{U}= \int \frac{d^2 k}{(2 \pi)^2} \frac{1}{2\xi_k+|\epsilon_B|},
\end{equation}
where $U \equiv -  g > 0$, and $\epsilon_B$ is the 2D two-body binding energy. In order to make contact with current experiments, it is convenient to relate $\epsilon_B$ to the three dimensional scattering length $a_s$. In the scattering of atoms confined in the axial direction by a harmonic potential with characteristic frequency $\omega_L$ they are related by~\cite{Petrov,Tempere}:

\begin{equation}
\label{EB}
|\epsilon_B|=\frac{C \omega_L}{\pi} exp\left( \sqrt{2 \pi} \frac{l_L}{a_s} \right),
\end{equation}
where $a_s$ is the 3D s-wave scattering length, $\omega_L=\sqrt{8 \pi^2 V_0/ (M \lambda^2)}$, $l_L=1/\sqrt{M\omega_L}$ is the axial ground state, and $C \approx 0.915$. $V_0$ is the amplitude of the periodic potential $V_0 \sin^2(2 \pi z/ \lambda)$ generated by two counter-propagating laser beans with length $\lambda$ parallel to the $z$-axis~\cite{Tempere}. It is worth to mention that if $l_L>a_s$, the relative wave function of molecules in quasi-2D Fermi gases explores higher transverse harmonic oscillator modes~\cite{baur}. In this case, i.e., in the limit of large binding energies $(|\epsilon_B|>\omega_L)$, effective corrections to the zero-range usual result which gives Eq.~(\ref{EB}) are required~\cite{baur}.

Thus we obtain an analytical expression for the grand canonical thermodynamic potential at $T=0$:

\begin{widetext}
\begin{eqnarray}
\bar\Omega &=&  \Omega(\Delta, \mu)+ \Theta_h \Omega(\Delta, h)\\
\nonumber
&=& \Delta^2\left[ \ln \left(\frac{\sqrt{\mu^2+\Delta^2}-\mu}{|\epsilon_B|}\right) -\frac{1}{2}  \right]-\mu\left(\sqrt{\mu^2+\Delta^2}+\mu \right)
- \Theta_h \left[2h \sqrt{h^2-\Delta^2}- \Delta^2 \ln \left(\frac{h+\sqrt{h^2-\Delta^2}}{h-\sqrt{h^2-\Delta^2}} \right) \right],
\label{LagTLM}
\end{eqnarray}
\end{widetext}
where $\bar\Omega \equiv \frac{4 \pi}{M}\Omega$, $\Theta_h \equiv \Theta(h^2-\Delta^2)$, $\mu=\frac{\mu_a+\mu_b}{2}$, $h=\frac{\mu_b-\mu_a}{2}$, and $\Theta(x)$ is the Heaviside step function, defined as $1$ if $x\geq0$, and $0$ if $x<0$. With these definitions, Eq.~(\ref{roots}) may be written as $k_{1,2}^2 = 2M(\mu \pm \sqrt{h^2-\Delta^2})$ and the condition for real $k_{1,2}^2$ is now $\Delta \leq h$.

\section{Phases of 2D Imbalanced Fermi Systems and Their Stabilities}
\label{P}

\subsection{Fixed Chemical Potentials}

In the grand canonical ensemble the chemical potentials of the two-components are held fixed. This happens when the system is connected to reservoirs of species $a$ and $b$ such that the particle densities are allowed to change. We analyze the ground state of this case for balanced and imbalanced configurations.

\subsubsection{Balanced Systems}

The gap and number equations are obtained by $\partial \Omega/ \partial \Delta =0$ and $n_{\alpha}=- \partial \Omega / \partial \mu_{\alpha}$, respectively. For the balanced system where $h=0$, which implies $\Theta_h=0$, we find

\begin{equation}
\sqrt{\mu^2+\Delta^2}-\mu = |\epsilon_B|,
\label{gap}
\end{equation}
and
\begin{equation}
\sqrt{\mu^2+\Delta^2}+ \mu = 2 \epsilon_F,
\label{mu}
\end{equation}
where the two dimensional Fermi energy is defined as $\epsilon_F = \frac{\pi n_T}{M}$, with $ n_T=n_a+n_b$, see Eq.~(\ref{eq2}). In the balanced configuration $n_{a}=n_{b} \equiv n = \frac{n_T}{2}$. Solving these two equations self-consistently we arrive at

\begin{equation}
\Delta_0=\sqrt{2\epsilon_F | \epsilon_B |},
\label{gap0}
\end{equation}
and

\begin{equation}
\label{mu0}
\mu_0=\epsilon_F - \frac{|\epsilon_B|}{2}.
\end{equation}

The value of the free energy at the minimum is

\begin{equation}
\Omega(h=0, \Delta=\Delta_0) \equiv \Omega_0 = - \kappa \left(\mu_0 + \frac{|\epsilon_B|}{2} \right)^2,
\label{min1}
\end{equation}
where $\kappa \equiv \frac{M}{2 \pi}$, whereas the energy of the balanced normal state is given by

\begin{equation}
\Omega(h=\Delta=0) \equiv \Omega_b^N = - \kappa \mu_0^2.
\label{min2}
\end{equation}
A direct comparison between Eqs.~(\ref{min1}) and (\ref{min2}) shows that the superfluid state is energetically preferable to the normal state for any $\epsilon_B \neq 0$. Since a two-body bound state exists even for an arbitrarily small attraction in 2D~\cite{Randeria}, the pairing instability will always happens in 2D balanced two-component Fermi systems at $T=0$.

\subsubsection{BCS-BEC Crossover in Balanced Systems}
\label{CrossoverBalanced}

The two well known~\cite{Randeria,Loktev} equations (\ref{gap0}) and (\ref{mu0}) reveal that the BCS-BEC crossover can be accessed by varying $\epsilon_B$ from weak to strong interaction regimes. The BCS state is reached in the weak attraction limit or high density ($\epsilon_F$), $|\epsilon_B| << \epsilon_F$, where $\mu_0 \cong \epsilon_F$ and the energy gap is found to be
\begin{equation}
\label{evalgap}
\Delta_0 = 2 \sqrt{\epsilon_F \rm{E}_{\Lambda}}~e^{-\frac{2 \pi}{M U}} << \epsilon_F,
\end{equation}
where $\rm{E}_{\Lambda} \equiv \frac{\Lambda^2}{2M}$, and $\Lambda$ is a momentum cutoff. To obtain Eq.~(\ref{evalgap}) from Eq.~(\ref{gap0}) we have made use of Eq.~(\ref{reg}). The opposite limit, of very strong attraction or low density, $|\epsilon_B| >> \epsilon_F$, results in the formation of composite bosons with $\mu_0 \cong - |\epsilon_B|/2$. Since the BCS and BEC regions are characterized by positive and negative chemical potentials, respectively, a clear distinction between these two regimes is the value of the binding energy at which $\mu_0$ changes sign i.e., $|\epsilon_B|=2\epsilon_F$.

\subsubsection{Imbalanced Systems}

We now turn our attention to the cases where $h \neq 0$. The free energy of the imbalanced normal state, $\Omega(h,\Delta=0) \equiv \Omega_i^N$, is found to be

\begin{eqnarray}
\Omega_i^N&=&\int_{k>k_2, k<k_1} \frac{d^2 k}{(2\pi)^2} (\epsilon_k^+ - |\epsilon_k^+|)\\
\nonumber
&+& \int_{k_1}^{k_2} \frac{d^2 k}{(2\pi)^2} \epsilon_k^b.
\end{eqnarray}
In the limit $\Delta \to 0$ we find $k_1(\Delta=0)=\sqrt{2M(\mu-h)}=k_F^a < k_F$ and $k_2(\Delta=0)=\sqrt{2M(\mu+h)}=k_F^b > k_F$, where $k_F=\sqrt{2M \mu}$. We are considering $\mu_{\alpha}$ and $h$ positive. Thus the equation above is written as:

\begin{eqnarray}
\Omega_i^N= \int_{0}^{k_F^a} \frac{d^2 k}{(2\pi)^2} \epsilon_k^a + \int_{0}^{k_F^b} \frac{d^2 k}{(2\pi)^2} \epsilon_k^b,
\end{eqnarray}
which gives, after the integration in $k$, the free energy of a (normal) two species gas of fermionic atoms in two dimensions

\begin{eqnarray}
\label{fn}
\Omega_i^N (\mu_a, \mu_b) &=& -\frac{\kappa}{2} [(\mu-h)^2 + (\mu+h)^2]
\nonumber
\\
&=& -\frac{\kappa}{2}[\mu_a^2 + \mu_b^2].
\end{eqnarray}

The gap and number equation now read

\begin{equation}
\label{gap2}
\ln \left (\frac{ \sqrt{\mu^2+\Delta^2}-\mu}{ |\epsilon_B|} \right) + \Theta_h  \ln \left(\frac{h+\sqrt{h^2-\Delta^2}}{h-\sqrt{h^2-\Delta^2}} \right)=0,
\end{equation}

\begin{equation}
\sqrt{\mu^2+\Delta^2}+ \mu = 2 \epsilon_F.
\label{n2}
\end{equation}
Note that for $h<\Delta$ Eq.~(\ref{gap2}) is reduced to Eq.~(\ref{gap}). Seeking now solutions where $h>\Delta$ we then have to solve

\begin{equation}
\label{gap3}
\frac{ \sqrt{\mu^2+\Delta^2}-\mu}{ |\epsilon_B|}  = \frac{h-\sqrt{h^2-\Delta^2}}{h+\sqrt{h^2-\Delta^2}}.
\end{equation}
Solving these equations self-consistently we find

\begin{equation}
\label{gaph}
\Delta_S(h)=\sqrt{\Delta_0(2h-\Delta_0)}.
\end{equation}
From the equation above we see that the ``Sarma state'' (as explained below) gap and the chemical potential imbalance exist in the ranges

\begin{equation}
\label{Deltah}
0 \leq \Delta_S(h) \leq \Delta_0,
\end{equation}

\begin{equation}
\label{rangeh}
\frac{\Delta_0}{2} \leq h \leq \Delta_0, 
\end{equation}
where the upper limit is imposed by the existence of real $k_{1,2}^2$.

\begin{figure}[htb]
  \vspace{0.5cm}
  \epsfig{figure=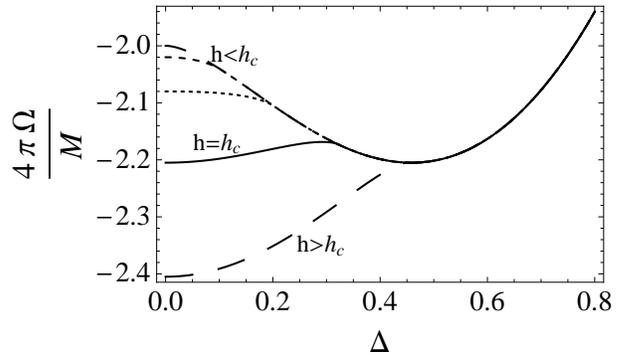,angle=0,width=8cm}
\caption[]{\label{omega} The thermodynamic potential $\Omega$ as a function of $\Delta$ for various chemical potential asymmetries. The top curve is for $h=0$. The imbalance is being increased from top to bottom and in all curves for $h < h_c$ the minimum is at $\Delta_0$. The curve for $h=h_c$ has two minima. After $h_c$, at which there is a quantum phase transition for the normal state, $\Omega(\Delta=0,h>h_c)<\Omega(\Delta_0,h \leq h_c)$.}
\end{figure}

From the graphical analysis of $\Omega$ as a function of $\Delta$ for various asymmetries, shown in Fig.~\ref{omega}, one sees that the first curve, from top to bottom, is the usual balanced system with its minimum at $\Delta_0$. Increasing the imbalance $h$ and keeping $\mu$ positive (to guarantee that we are in the BCS regime), the minimum is still located at $\Delta_0$ up to a maximum or critical imbalance $h_c$, after which there is a quantum phase transition to the normal state with $\Delta=0$. $h_c$ is find through the equality $\Omega_0=\Omega_i^N$, which yields:

\begin{equation}
h_c^2= \mu_0 |\epsilon_B| + \left( \frac{\epsilon_B}{2} \right)^2,
\end{equation}
from which one easily finds plugging in the equation above $\mu_0$ from Eq.~(\ref{mu0}) in the BCS limit ($|\epsilon_B| << \epsilon_F$): 

\begin{equation}
h_c=\frac{\Delta_0}{\sqrt{2}},
\end{equation}
which is the same ${\rm 3D}$ result known as the Chandrasekhar-Clogston limit of superfluidity~\cite{Chandrasekhar,Clogston}.

We have seen above that, as happens in 3D~\cite{Caldas1,Caldas2,JSTAT1}, the BCS phase turns from stable, while $h < h_c$, to metastable, when $h > h_c$, in which case the normal phase is stable. Besides these two phases, there is an unstable phase, known as Sarma state, corresponding to a local maximum of $\Omega$ versus $\Delta$, that is located between the BCS minimum at $\Delta_0$ and the normal phase with $\Delta=0$.

\subsection{Fixed Number Densities}

In the previous subsection we discussed the stability of several phases with fixed chemical potentials $\mu_a$ and $\mu_b$. We investigate now the situations where the number densities $n_a$ and $n_b$ are fixed, since it is appropriate to describe trapped gases in current experiments. 

It is convenient to introduce now the ``magnetization'' defined as $m=-\frac{d \Omega}{d h}=n_b-n_a$, which will be needed to describe the energy of the Sarma phase in terms of the number densities. The total derivative of $\Omega=\Omega(\mu_a,\mu_b,\Delta)$ is written as

\begin{equation}
d \Omega = \frac{\partial \Omega}{\partial \mu_a} d \mu_a + \frac{\partial \Omega}{\partial \mu_b} d \mu_b + \frac{\partial \Omega}{\partial \Delta} d \Delta.
\label{tot}
\end{equation}
Since the densities have to evaluated at the minimum of $\Omega$, i.e., at $\Delta_0$, the last term vanishes. The particle densities are given by $n_{\alpha}=- \frac{\partial \Omega}{\partial \mu_{\alpha}}$. Then, using $\mu_a=\mu-h$ and $\mu_b=\mu+h$ in Eq.~(\ref{tot}) we find

\begin{equation}
m=n_{b}-n_{a}=-\frac{d \Omega}{dh},
\label{eq1}
\end{equation}
and

\begin{equation}
n_T=n_{b}+n_{a}=-\frac{d \Omega}{d \mu}.
\label{eq2}
\end{equation}

\noindent Then the magnetization can be written as
 
\begin{equation}
m=\frac{M}{\pi} \sqrt{h^2-\Delta^2}.
\label{mag}
\end{equation}
Solving equations (\ref{n2}), (\ref{gap3}) and (\ref{mag}) self-consistently, we find the gap $\Delta_{S}(m)$ and the ``average'' chemical potential $\mu_S$ of the Sarma phase

\begin{equation}
\Delta_{S}(m)=\sqrt{\Delta_0^2-\frac{2 \pi \Delta_0}{M}m},
\label{gapS}
\end{equation}
where $\Delta_0$, the gap parameter of the balanced system, is given by Eq.~(\ref{gap0}), and

\begin{equation}
\mu_S=\epsilon_F - \frac{{\Delta_S}^2}{4 \epsilon_F}.
\label{mu_S}
\end{equation}
Note that the equations above reduce to Eqs.~(\ref{gap0}) and (\ref{mu0}), respectively, since in the limit $m \to 0$, $\Delta_S \to \Delta_0$, as it should be. From Eq.~(\ref{gapS}) it is easy to obtain the windows for the Sarma gap and $m$:

\begin{equation}
0 \leq \Delta_{S}(m) \leq \Delta_0,
\label{gapS2}
\end{equation}

\begin{equation}
0 \leq m \leq m_{max},
\label{m_max}
\end{equation}
where $m_{max}=\frac{M \Delta_0}{2 \pi}$ is the maximum value for the density imbalance in the Sarma phase, which is easily obtained from Eq.~(\ref{gapS}).

The expression for the energy $E$ (not the free energy or thermodynamic potential) of the homogeneous Sarma phase is obtained as:

\begin{equation}
E(n_T,m,\Delta)= \Omega(\mu,h,\Delta) + \mu_a n_a + \mu_b n_b.
\label{eq3}
\end{equation}
To write $E=E(n_T,m)$, we express $\mu_a n_a + \mu_b n_b=\mu n_T + h m$, with $\mu=\mu(n_T)$, $h=h(m)$ to find:

\begin{widetext}
\begin{eqnarray}
\label{eq33}
E(n_T,m,\Delta) &\equiv& E^S(n_a,n_b)= \Omega(n_T,m,\Delta) + \mu n_T + h m 
\nonumber
\\
&=& \frac{M}{4 \pi} \left\{ \Delta_S^2 \left[ \ln \left( \frac{\Delta_S^2}{\Delta_0^2} \right) -\frac{1}{2} \right] - 2\epsilon_F \left( \epsilon_F - \frac{\Delta_S^2}{4\epsilon_F} \right) -2 \sqrt{\bar m^2 + \Delta_S^2} ~\bar m + \Delta_S^2 \ln \left( \frac{\sqrt{\bar m^2 + \Delta_S^2} + \bar m}{\sqrt{\bar m^2 + \Delta_S^2} - \bar m} \right) \right\}
\nonumber
\\
&+& \left( \epsilon_F - \frac{\Delta_S^2}{4\epsilon_F} \right) n_T + \sqrt{\bar m^2 + \Delta_S^2}~m ,
\end{eqnarray}
\end{widetext}
where $\bar m \equiv \frac{\pi m}{M}$.

To investigate which one is the ground state of an imbalanced system with different densities $n_a$ and $n_b$, we have to see which energy of the following states we are considering is smaller, the normal $E^N(n_a, n_b)$, see Eq.~(\ref{EN}) below, the homogeneous Sarma $E^S(n_a,n_b)$, or the phase separation (PS) $E^{PS}(n_a, n_b)$ state. The PS is an inhomogeneous phase where given $n_a$ and $n_b$ densities in a trap, a fraction $1-x$ of the 2D (real) space is in the BCS phase where both species have a common density $n_a^{BCS}= n_b^{BCS} = n$ and the rest of particles are in the normal phase occupying the fraction $x$ around \footnote{A surface energy is related to the interface between the normal and superfluid phases~\cite{SE1,SE2,SE3,SE4}. The surface tension associated with this interface has been interpreted as responsible for deformations in highly elongated small samples~\cite{Exp6,Exp66}. These results are not consistent with the observations reported in Ref.~\cite{Exp7}. However, recent Rice experiments~\cite{NoST} find that their earlier results are indicated to be due to metastable states rather than surface tension. Thus we neglect the surface energy in the present analysis.} the BCS core, with densities $\tilde n_a$ and $\tilde n_b$. The most favored composition minimizes:

\begin{equation}
\label{PS}
E^{PS}=Min_{x, \tilde n}[(1-x)E(n)+ xE^N(\tilde n_a, \tilde n_b)],
\end{equation}
where the space fraction is obviously limited to the range $0 \leq x \leq 1$. The energy of the balanced superfluid phase is found as

\begin{eqnarray}
\label{BCS(n)}
E(n) &=& \Omega(\Delta, h=0, \mu) + \mu n_T
\nonumber
\\
&=& \frac{M}{4 \pi} \left\{ \Delta^2 \left[ \ln \left( \frac{\Delta^2}{\Delta_0^2} \right) -1 \right] + 2 \epsilon_F^2 \right\}
\nonumber
\\
&=&  \frac{M \Delta^2}{4 \pi}  \left[ \ln \left( \frac{\Delta^2}{\Delta_0^2} \right) -1 \right] + \frac{\pi (2n)^2}{2M}.
\end{eqnarray}
Since $E(n)$ which enters $E^{PS}$ has to be written at its minimum, the equation above turns out to be:

\begin{eqnarray}
\label{BCS(n)2}
E(n) &=&  \frac{\pi {n_T}^2}{2M} -\frac{M \Delta_0^2(n)}{4 \pi}
\nonumber
\\
&=& \frac{2 \pi}{M} n^2 - |\epsilon_B|  n,
\end{eqnarray}
where the gap parameter of the BCS phase is given by Eq.~(\ref{gap0}) and written as $\Delta_0(n)=2\sqrt{\frac{\pi}{M} | \epsilon_B | n}$.

The energy of the normal state entering in Eq.~(\ref{PS}) is easy to find from Eq.~(\ref{fn}) as

\begin{eqnarray}
\label{EN}
E^N(n_a, n_b)&=& \Omega^N(n_a, n_b) + \mu_a n_b + \mu_a n_b
\nonumber
\\
 &=& \frac{1}{2 \kappa} (n_a^2 + n_b^2).
\end{eqnarray}
Note that since $n_a= \frac{n_T - m}{2}$ and $n_b= \frac{n_T + m}{2}$, the equation above for $m=0$ gives $E^N=\frac{\pi n_T^2}{2 M}$ agreeing with the result of Eq.~(\ref{BCS(n)}) in the limit $\Delta \to 0$.

As we mentioned before, the $n_a$ and $n_b$ particle densities are accommodated in the trap as

\begin{eqnarray}
\label{n}
n_a &=& x \tilde n_a + (1-x) \tilde n,
\nonumber
\\
n_b &=& x \tilde n_b + (1-x) \tilde n.
\end{eqnarray}
Then we rewrite Eq.~(\ref{EN}) as

\begin{eqnarray}
\label{EN2}
&& E^N(\tilde n_a, \tilde n_b) =
\nonumber
\\
&& \frac{1}{2 \kappa} \left[ \left(\frac{n_a - (1-x) \tilde n}{x} \right)^2 + \left(\frac{n_b - (1-x) \tilde n}{x} \right)^2 \right],
\end{eqnarray}
 
It is interesting to note that Eqs.~(\ref{BCS(n)2}) and (\ref{EN}) are easily obtained from Eq.~(\ref{eq33}) in the limits $m \to 0$ and $m \to m_{max}$, respectively.

In the minimization of $E^{PS}$ we shall find  $x \approx 0$ and $\tilde n \approx n_a$ ($~\rm{or}$ $ n_b$) if $n_b \approx n_a$, which means that the entire system is in the superfluid phase. In the other limit, if $n_b >> n_a$, the system will be in the normal state, so we shall find in the minimization of $E^{PS}$ $x \approx 1$ and $\tilde n \approx 0$. In the intermediate cases the system will phase separate, i.e., if the imbalance is less than the critical value, as experiments have shown for 3D gases.
\newline
\newline

\subsubsection{Phase Competition}

There are some limiting cases where the comparison between the phase separation and Sarma states can be done analytically and exactly. The first one is obtained when $m=0$, giving $E^S(n)=E^{PS}(n)=E^{BCS}(n)$. The second case happens when $m=m_{max}$, which reduces the Sarma phase to the normal phase with $\Delta=0$. This is easily seen from Eqs.~(\ref{m_max}) and (\ref{gapS2}). Then we define $\Delta E = E^{PS}-E^N$, which is written as

\begin{eqnarray}
\label{DeltaE}
\Delta E &=& \left[\frac{\tilde n^2}{\kappa}-E(\tilde n)\right](x-1)
\nonumber
\\
&+& \frac{1}{2 \kappa} \left[ (n_T -\tilde n)^2 -2 n_a n_b +\tilde n^2 \right]\left(\frac{1}{x} -1 \right).
\end{eqnarray}
Minimization with respect to $x$ yields:

\begin{equation}
\label{xmin1}
x_{min}=\sqrt{\frac{(n_T-\tilde n)^2 -2n_an_b +\tilde n^2}{2(\tilde n^2 - \kappa E(\tilde n))}}.
\end{equation}

\begin{figure}[htb]
  \vspace{0.5cm}
\epsfysize=5.5cm
  \epsfig{figure=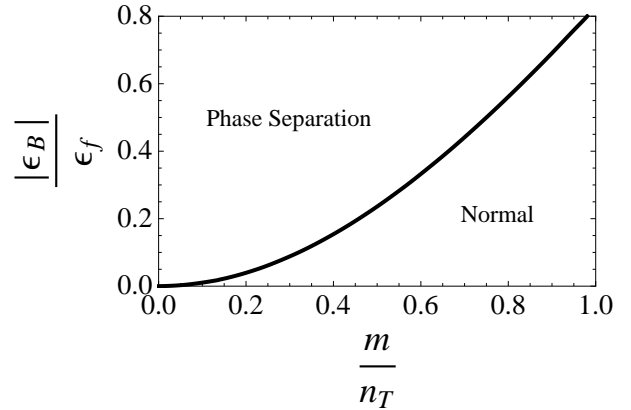,angle=0,width=8cm}
\caption[]{\label{pd} The phase diagram of a zero temperature imbalanced gas of fermionic atoms in 2D. The vertical axis $|\epsilon_B|/\epsilon_F$ with $m/n_T=0$, is the (balanced) BCS phase, whereas the horizontal line $|\epsilon_B|/\epsilon_F=0$ represents the (imbalanced) normal phase. Note that for each value of $|\epsilon_B|/\epsilon_F$ phase separation is the ground state up to $m_c/n_T$. Above this value phase separation is not energetically favorable, and the normal phase has lower energy.}
\end{figure}

An upper bound on $\Delta E$ can be obtained by setting the density of the BCS component of the PS state $\tilde n=n_a$ and writing $n_b= n_a + m$, where for low imbalance, $m <~n_a, n_b$, Eq.~(\ref{xmin1}) gives:

\begin{equation}
\label{xmin}
 x_{min}=  \frac{m}{ \sqrt{ \frac{M^2 \Delta_0^2(n_a)}{4 \pi^2} + \frac{3}{2} n_a^2 }},
\end{equation}
and

\begin{eqnarray}
\label{deltaE}
\Delta E(x_{min}) &=& - \left( \frac{M \Delta_0^2(n_a)}{4 \pi} + \frac{3 \pi}{2 M} n_a^2 \right) 
\nonumber
\\
&\times& (1-x_{min})^2 < 0.
\end{eqnarray}
Unlike the three-dimensional case where the result obtained is an approximation, valid only for small asymmetries $m << n_a, n_b$, where terms of order ${m}^3/{na}^{4/3}$ have been neglected~\cite{Caldas1,Caldas2}, the result above is exact and valid for any $m < n_a, n_b$. These results agree with the numerical analysis made to construct the phase diagram depicted in Fig.~(\ref{pd}).

Let us now show the results of the minimization of Eq~(\ref{PS}) with respect to $x$ and $n$ for a given $n_a$ and $n_b$. The results are shown in Fig.~(\ref{pd}).

\begin{figure}[htb]
 \centerline{ \epsfig{file=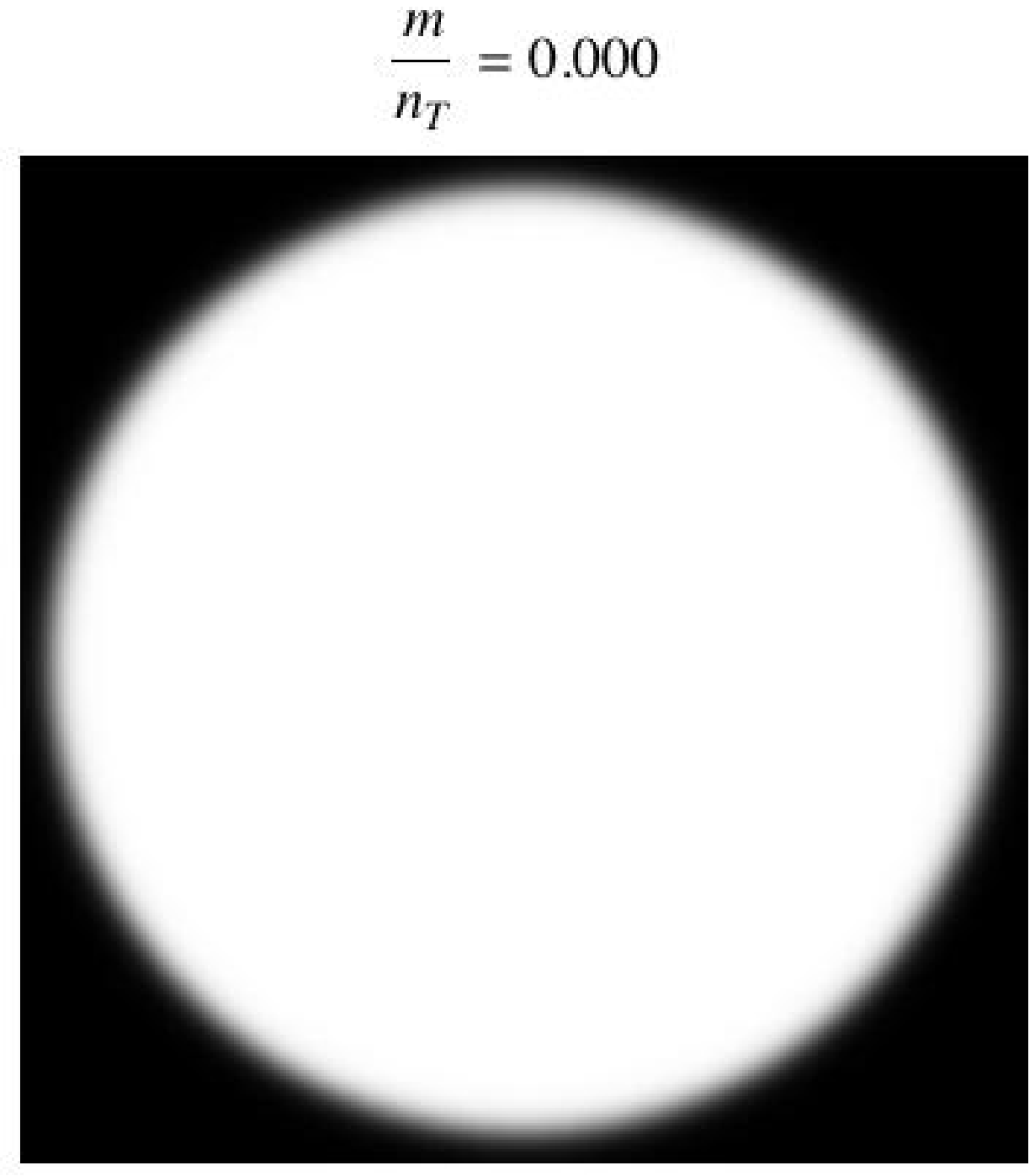,scale=0.30,angle=0} 
   \epsfig{file=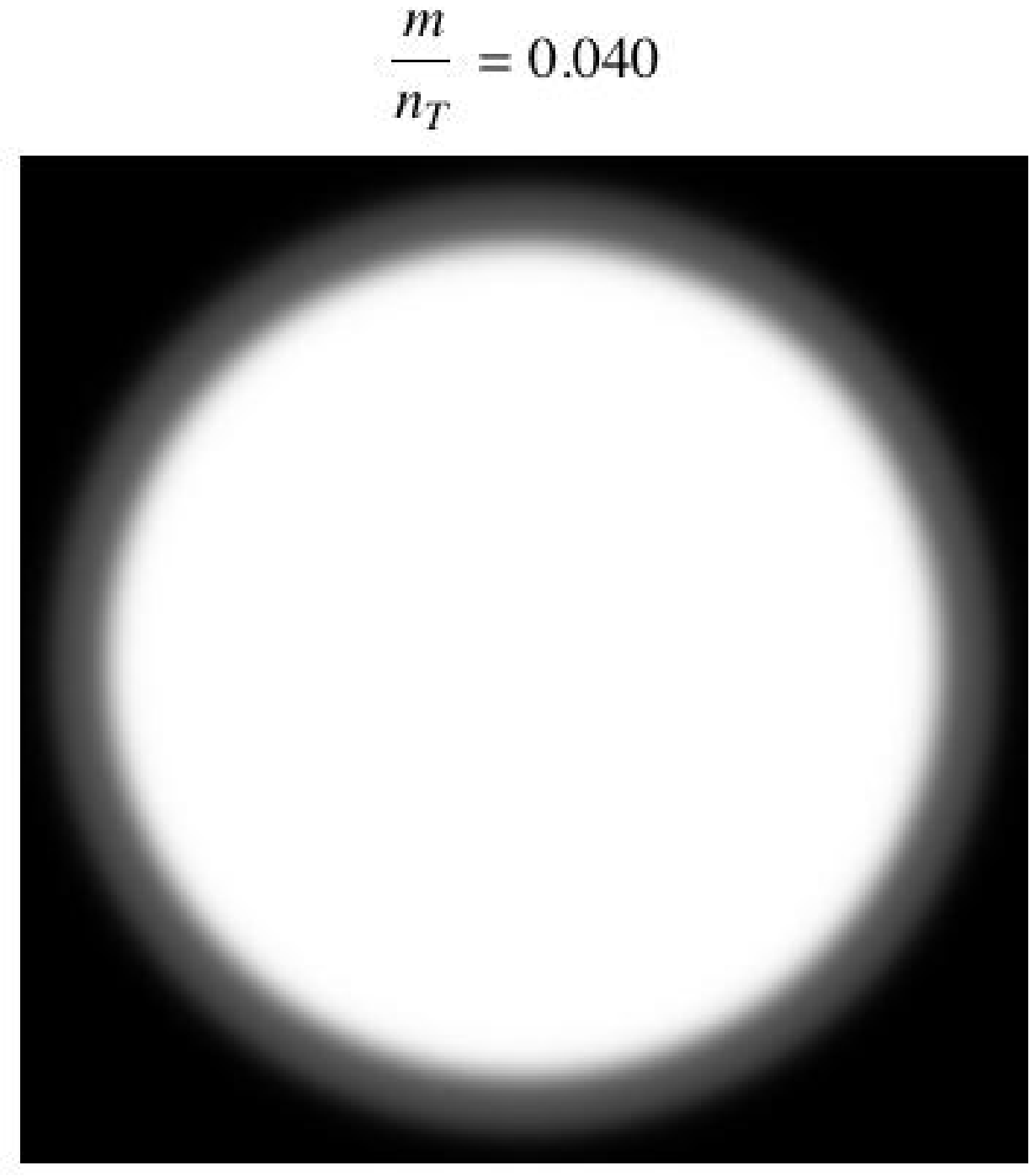,scale=0.30,angle=0}}
  \centerline{ \epsfig{file=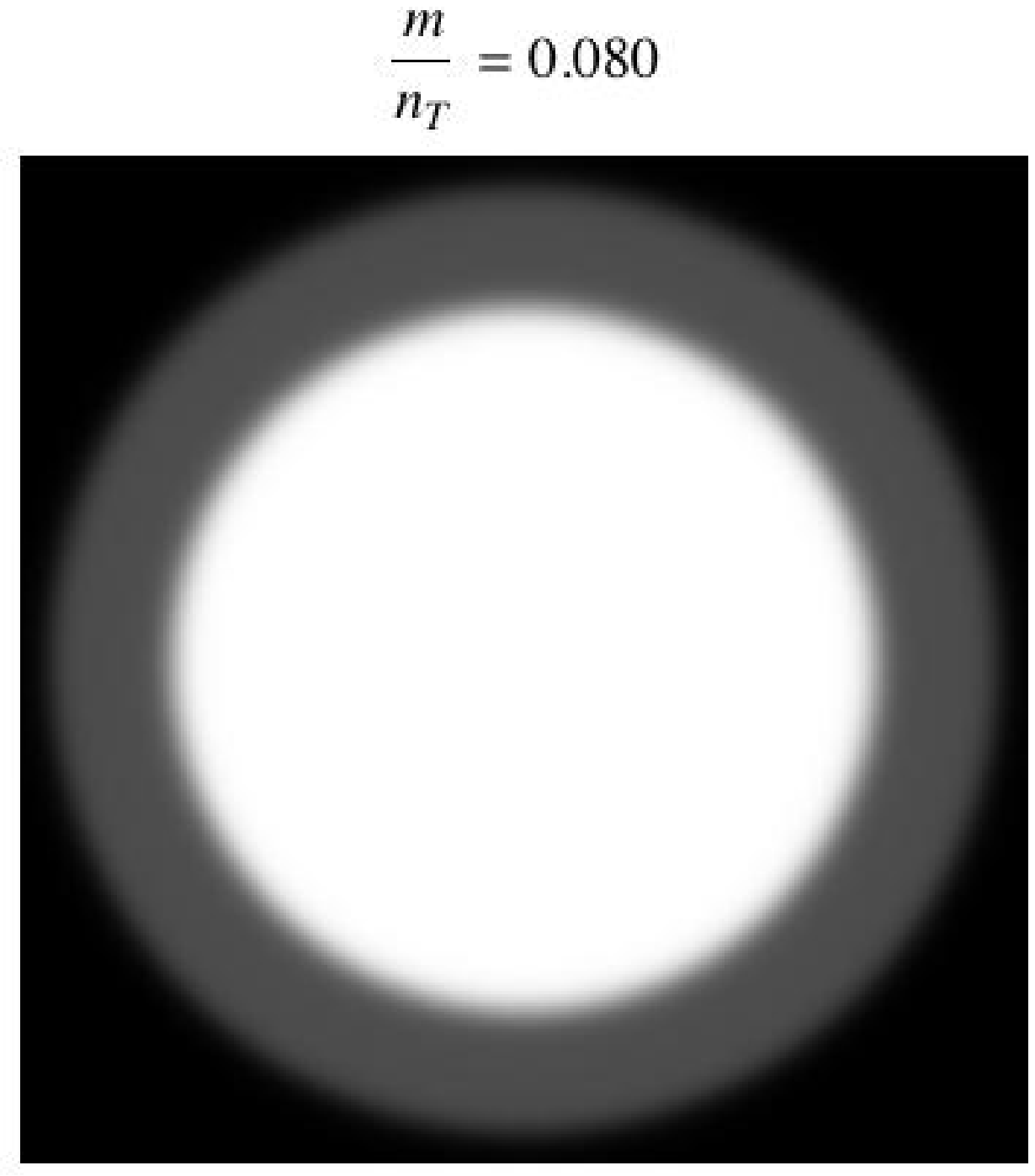,scale=0.30,angle=0} 
   \epsfig{file=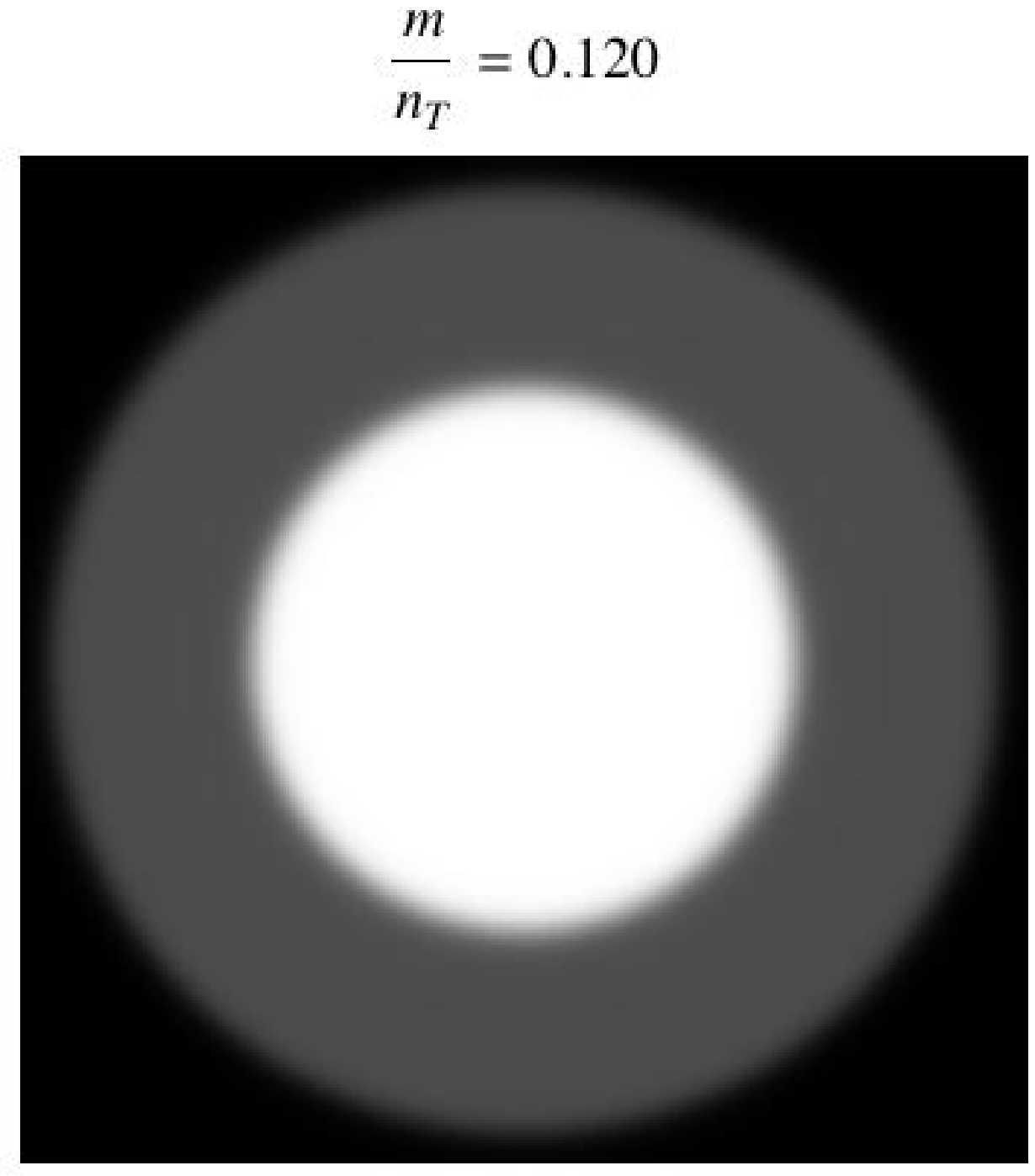,scale=0.30,angle=0}}
   \centerline{ \epsfig{file=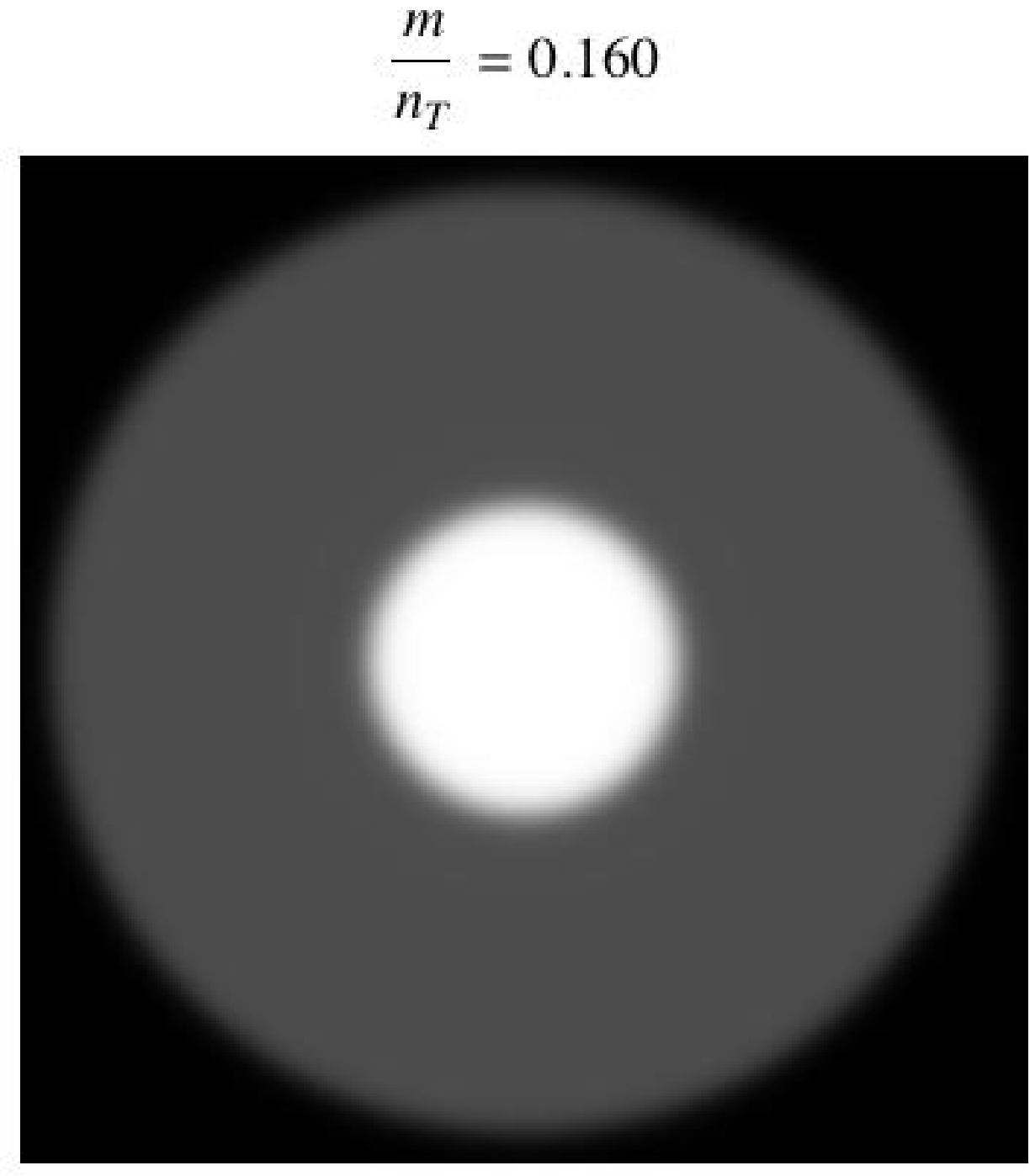,scale=0.30,angle=0} 
   \epsfig{file=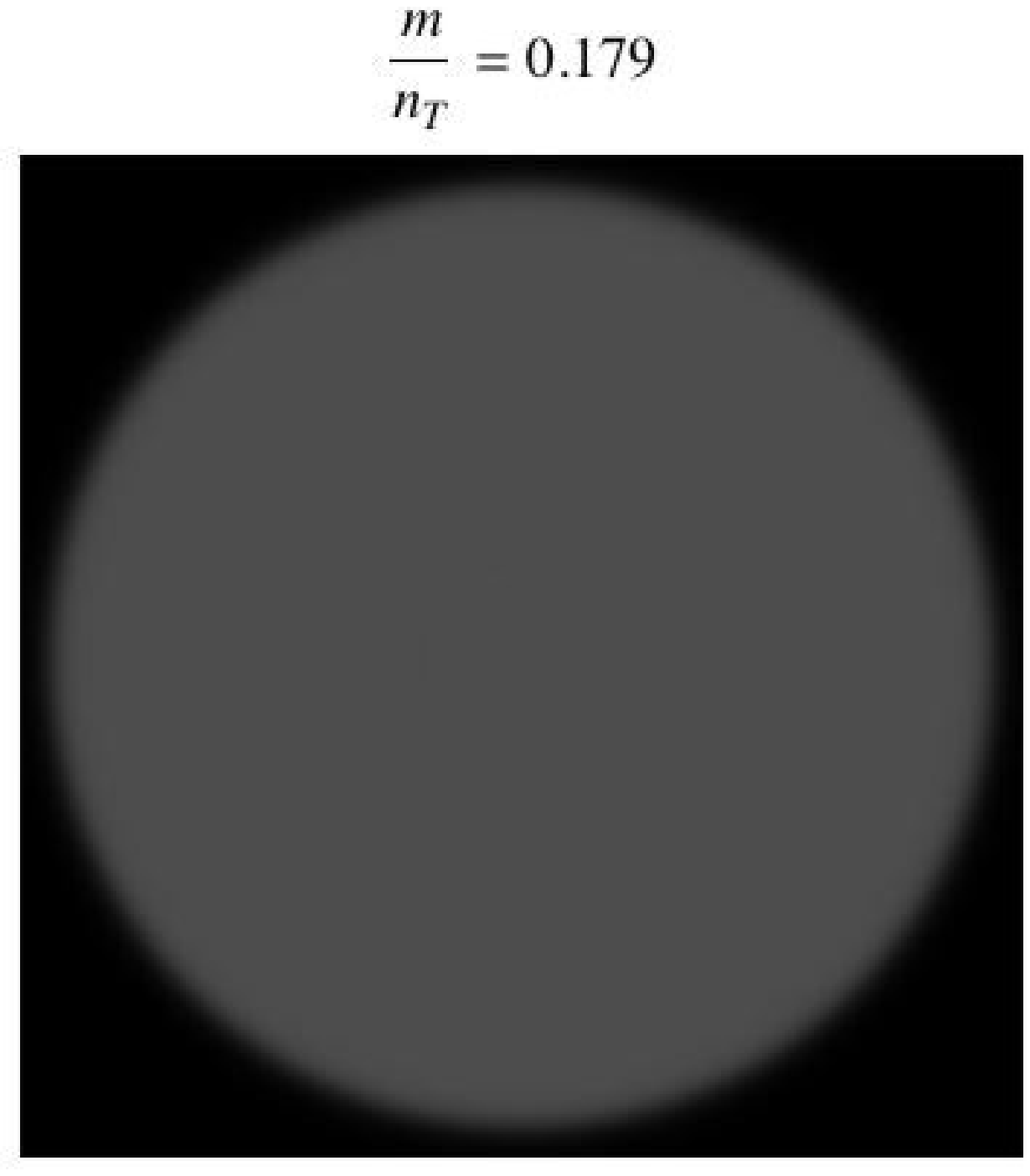,scale=0.30,angle=0}}
\caption[]{\label{cartoon} Evolution of the phases of an imbalanced gas with fixed $|\epsilon_B|/\epsilon_F=0.0318$ as a function of increasing $m/n_T$. The first picture, with $m/n_T=0$ represent the BCS phase. Increasing $m/n_T$ the system phase separates (pictures 2 to 5), and increasing the imbalance even more, the system finally goes to the normal phase at the critical imbalance $m_c/n_T=0.179$, as seen in the last picture.}
\end{figure}

We show in Fig.~(\ref{cartoon}) an illustration of what may be found experimentally in an imbalanced $2D$ gas of femionic atoms. The first picture represents the balanced system with $m=0$. The following pictures represent a set of experiments with fixed $\epsilon_B/\epsilon_F=0.0318$ with increasing imbalance. From the second picture to the fifth the system phase separates and the sixth picture is for the critical imbalance $m_c$ representing the normal phase.

\subsubsection{BCS-BEC Crossover in Imbalanced Systems}

Now we investigate the BCS-BEC crossover in an imbalanced gas of fermionic 
atoms. As we have seen, there are three possible stable phases with fixed number 
densities. For zero imbalance between the densities, the only phase present is BCS 
(for any value of the interaction between the spin up and down fermionic atoms) and 
the crossover is that discussed in Subsection~\ref{CrossoverBalanced}, which is 
governed only by $\epsilon_B$. For imbalances where phase separation is possible, 
the BCS-BEC crossover can be realized in the superfluid core of the separated phases, 
again varying $\epsilon_B$ in $\mu_0=\frac{2 \pi \tilde n}{M} - 
\frac{|\epsilon_B|}{2}$, where $\tilde n$ is the BCS density that minimizes the PS 
formation, as described above Eq.~(\ref{PS}). For large enough asymmetries, phase 
separation is not supported and the entire system is in the normal phase with no 
possibility of pairing formation and the crossover to the BEC regime. Then, when $m 
\neq 0$ the BCS-BEC crossover can not be accessed only by varying the biding energy 
since the imbalance plays an important role now.

\section{Bound States in the Normal Phase}
\label{bs}

In this section we investigate the necessary conditions for the formation of bound states in the normal phase. We expand the action up to second order in the order parameter $|\Delta_{\vec{q}}|$, and obtain

\begin{equation}
 S_{\text{eff}}=\sum_{\vec{q},\omega_0}\alpha({|\vec{q}|}, \omega_0) |\Delta_{\vec{q}}|^{2}+
\mathcal{O}\left(|\Delta|^4\right)\,,
\end{equation}
where $\alpha(|{\vec{q}}|,\omega_0)=\frac{1}{g}-\chi({|\vec{q}|}, \omega_0)$, and $\chi(|{\vec{q}}|, \omega_0)$ is the generalized pair susceptibility:

\begin{eqnarray}
\label{chi1-1}
\chi(|{\vec{q}}|, \omega_0)=
\sum_{\vec{k}} \frac{1-n(\xi_{\vec{k}-\vec{q}/2,\uparrow})-n(\xi_{\vec{k}+\vec{q}/2,\downarrow})}{\xi_{\vec{k}-\vec{q}/2,\uparrow}+\xi_{\vec{k}+\vec{q}/2,\downarrow}-\omega_0}.
\end{eqnarray}
Evaluation of the equation above is straightforward. At zero temperature we find that $\chi(|{\vec{q}}|, \omega_0)$ is given by:

\begin{widetext}
\begin{eqnarray}
\label{chi1-2}
\chi(|{\vec{q}}|, \omega_0) &=& N(0) \left[\ln\left(\frac{2 \omega_c}{h} \right) - \ln \left(1 - \frac{\omega_0}{2h}+ \sqrt{ \left(1 - \frac{\omega_0}{2h} \right)^2 -{\bar q}^2} \right) \right],\\
\nonumber
&&{\rm for}~ \bar q \leq 1 - \frac{\omega_0}{2h},
\end{eqnarray}
\end{widetext}
where $\omega_c$ is an energy cutoff, $N(0)=\frac{m}{2 \pi}$ is the density of states at the Fermi level, $\bar q \equiv \frac{v_F |q|}{2h}$, and $v_F$ is the Fermi velocity. 

The spectrum of bound states is given by $1/g-\chi(|{\vec{q}}|=0, \omega_0)=0$, which corresponds to the Thouless criterion $1-g \chi=0$, signalizing the divergence of the T-matrix~\cite{Loktev}. Using Eq.~(\ref{reg}) to trade the cutoff in favor of the  2D two-body binding energy $\epsilon_B$, we obtain the following real (and unique) equation for the energies of these states in the normal phase at zero $T$:

\begin{equation}
\label{bond}
\omega_0=2h-(2\mu+|\epsilon_B|).
\end{equation}
In the balanced normal phase with $h=0$ we obtain $\omega_0=-(2\mu+|\epsilon_B|)$ that, for $\mu>0$, corresponds to the energy of Cooper pairs, agreeing with previous work \cite{Loktev}. In the imbalanced case, bound states exist for $2\mu+|\epsilon_B|> 2h$.

\section{Comparison With Some Previous Zero Temperature Results}
\label{comp}

In this paper using the mean-field approximation we have studied the zero temperature ground state of a 2D imbalanced gas of fermonic atoms. We performed our calculations in the Grand Canonical and Canonical ensembles, considering chemical potential and population imbalance, respectively. It is very interesting to note that our results corroborate with solid results obtained in the recent literature. In particular, the result of our Fig.~\ref{pd} is equivalent to the diagram (bottom panel of Fig.~2) in the work of He and Zhuang~\cite{Zhuang}. In this work we have shown that if you change the ensemble the results are the same, as expected, since it is known that the physics can not depend on the ensemble used, and we have verified this in detail. 

Similar studies and results can be found in other very interesting and recent papers~\cite{Tempere,Simons}, and we have obtained the same results of these papers where the order of the phase transition from BCS to normal state at a critical chemical potential imbalanced is first order. Results II mentioned in our introduction also agree with the results found in~\cite{Simons}.

In Ref.~\cite{Du} the mean-field zero $T$ phase diagram of an imbalanced Fermi gas was investigated when there is also mass imbalance. Although they did not consider phase separation, they studied the influence of population and mass imbalance on the different Fermi surfaces topologies, associated to stable and unstable phases.

Spin-orbit coupling (SOC) $\lambda$ has been taken into account in Ref.~\cite{Wan} in a imbalanced 2D Fermi gas, and it was found that for large values of the SOC there is a topological phase transition from phase separation to a nontrivial superfluid phase. Our phase diagram of Fig. 2 is in complete agreement with the ($\lambda=0$ i.e., the vertical axes of the) zero $T$ mean-field phase diagrams of Fig. 1 shown in Ref.~\cite{Wan}.

\section{Conclusions}
\label{conc}

We have investigated superfluidity in two-dimensional imbalanced Fermi gases. With 
exact expressions, we have constructed the zero temperature mean-field phase diagram 
of these systems. We show that in the grand canonical ensemble, BCS is the ground state 
up to a critical chemical potential imbalance $h_c$ at which there is a first-order phase 
transition to the normal state. In the canonical ensemble, relevant to current experiments, 
we also show that, as found theoretically and experimentally in 3D, for a fixed two-body 
binding energy $\epsilon_B$, as the imbalance $m$ is increased from zero the stable states are 
BCS, phase separation and normal phase. 

Regarding the BCS-BEC crossover, in balanced systems it is governed only by $\epsilon_B$. However, for imbalanced systems, the BCS-BEC crossover will 
depend on $\epsilon_B$ and the imbalance $m$. For a fix value of $\epsilon_B$, the crossover is possible 
only for asymmetries supported by phase separation. Above a critical value of the 
number densities imbalance $m_c$, phase separation is destroyed in favor of the normal 
phase, as well as the path to BEC regime. Thus, for a given $m<m_c$ such that the system 
phase separates into BCS and normal phases, the BEC regime can be accessed by varying 
$\epsilon_B$ from the weak to strong couplings. A smaller and smaller circular core shall be 
visibly apparent, as $\epsilon_B$ is increased, signalizing the condensation of the Cooper pairs. 
This is a direct prediction of our results that could be tested experimentally.

We have investigated the conditions for the existence of bound states in the normal phase. As we have seen, since a two-body bound state exists even for an arbitrarily small attraction in 2D, the normal balanced state at $T=0$ corresponds to the superfluid case. In the imbalanced normal phase, bound states are likely to be formed provided $2\mu+|\epsilon_B|> 2h$.

As we mentioned in the introduction Section, the Fermi polaron problem appears in a very large imbalance $m>m_c$ or, in the unities of Fig.~\ref{pd}, $m/n_T \approx 1$. In this situation, the minority species $a$ correspond to impurities immersed in the majority species $b$ Fermi liquid. As the interaction $|\epsilon_B|/\epsilon_F$ is increased, the impurities $a$ surround themselves with a localized cloud of environment $b$ atoms, forming Fermi polarons~\cite{MartinComment}. A very interesting phenomena that could also be verified experimentally appears when the value of $|\epsilon_B|/\epsilon_F$ is fixed and the number of impurities is increased, i.e., one starts with an imbalance $m \approx n_T$ and end up with $m<m_c$. In this way, we expect a transition from polarons to phase separation configuration, since evidence of quasiparticle interactions have been found as the concentration of $a$ species were increased in the Fermi sea of $b$ atoms~\cite{nature_khol}.

Finally, we hope that the analytical expressions we have obtained here could be useful for other works studying 2D imbalanced Fermi gases.

\section{Acknowledgments}

We thank Drs. M. W. Zwierlein, L. H. C. M. Nunes, J. Devreese and M. Continentino for enlightening discussions. We also thank Dr. R. G. Hulet for a critical reading of the manuscript and for valuable suggestions. H. C. and A. L. M. are partially supported by CNPq. The authors also acknowledge partial support from FAPEMIG.


\end{document}